\documentclass[a4paper]{jpconf}
\usepackage{graphicx}
 \usepackage{setspace}

\def\BBz{0$\nu\beta\beta$}
\def\BBt{2$\nu\beta\beta$}

\def\mj{M{\sc ajo\-ra\-na}}
\def\dem{D{\sc e\-mon\-strat\-or}}

\def\QBB{Q$_{\beta\beta}$}

\def\ge{$^{76}$Ge}
\def\xe{$^{136}$Xe}
\def\te{$^{130}$Te}

\begin{document}
\title{The \mj\ \dem: A Search for Neutrinoless Double-beta Decay of \ge}
\author{W.~Xu$^{1}$,
N.~Abgrall$^{2}$, 
F.~T.~Avignone III$^{3,4}$, 
A.~S.~Barabash$^{5}$, 
F.~E.~Bertrand$^{4}$, 
V.~Brudanin$^{6}$, 
M.~Busch$^{7,8}$, 
M.~Buuck$^{9}$,
D.~Byram$^{10}$,
A.S.~Caldwell$^{11}$,
Y-D.~Chan$^{2}$, 
C.~D.~Christofferson$^{11}$, 
C.~Cuesta$^{9}$, 
J.~A.~Detwiler$^{9}$, 
Yu.~Efremenko$^{12}$, 
H.~Ejiri$^{13}$, 
S.~R.~Elliott$^{1}$, 
A.~Galindo-Uribarri$^{4}$, 
G.~K.~Giovanetti$^{14,8}$, 
J.~Goett$^{1}$,
M.~P.~Green$^{4}$, 
J.~Gruszko$^{9}$,
I.~Guinn$^{9}$,
V.~E.~Guiseppe$^{3}$, 
R.~Henning$^{14,8}$, 
E.~W.~Hoppe$^{15}$, 
S.~Howard$^{11}$, 
M.~A.~Howe$^{14,8}$, 
B.~R.~Jasinski$^{10}$, 
K.~J.~Keeter$^{16}$, 
M.~F.~Kidd$^{17}$, 
S.~I.~Konovalov$^{5}$, 
R.~T.~Kouzes$^{15}$, 
B.~D.~LaFerriere$^{15}$, 
J.~Leon$^{9}$, 
J.~MacMullin$^{14,8}$, 
R.~D.~Martin$^{10}$, 
S.~J.~Meijer$^{14,8}$,
S.~Mertens$^{2}$,
J.~L.~Orrell$^{15}$, 
C.~O'Shaughnessy$^{14,8}$,
N.~R.~Overman$^{15}$, 
A.~W.~P.~Poon$^{2}$, 
D.~C.~Radford$^{4}$, 
J.~Rager$^{14,8}$,
K.~Rielage$^{1}$, 
R.~G.~H.~Robertson$^{9}$, 
E.~Romero-Romero$^{12,4}$,
M.~C.~Ronquest$^{1}$, 
B.~Shanks$^{14,8}$,
M.~Shirchenko$^{6}$, 
N.~Snyder$^{10}$,
A.~M.~Suriano$^{11}$,
D.~Tedeschi$^{3}$,
J.~E.~Trimble$^{14,8}$,
R.~L.~Varner$^{4}$, 
S.~Vasilyev$^{6}$,
K.~Vetter$^{2}$ \footnote[18]{Alternate Address: Department of Nuclear Engineering, University of California,
Berkeley, CA, USA},
K.~Vorren$^{14,8}$, 
B.~R.~White$^{4}$,
J.~F.~Wilkerson$^{14,8,4}$, 
C.~Wiseman$^{3}$,
E.~Yakushev$^{6}$, 
C-H.~Yu$^{4}$,
and V.~Yumatov$^{5}$\\The \mj\ Collaboration}

\address{$^{1}$Los Alamos National Laboratory, Los Alamos, NM, USA}
\address{$^{2}$Nuclear Science Division, Lawrence Berkeley National Laboratory, Berkeley, CA, USA}
\address{$^{3}$Department of Physics and Astronomy, University of South Carolina, Columbia, SC, USA}
\address{$^{4}$Oak Ridge National Laboratory, Oak Ridge, TN, USA}
\address{$^{5}$Institute for Theoretical and Experimental Physics, Moscow, Russia}
\address{$^{6}$Joint Institute for Nuclear Research, Dubna, Russia}
\address{$^{7}$Department of Physics, Duke University, Durham, NC, USA}
\address{$^{8}$Triangle Universities Nuclear Laboratory, Durham, NC, USA}
\address{$^{9}$Center for Experimental Nuclear Physics and Astrophysics and \\
             Department of Physics, University of Washington, Seattle, WA, USA}
\address{$^{10}$Department of Physics, University of South Dakota, Vermillion, SD, USA}
\address{$^{11}$South Dakota School of Mines and Technology, Rapid City, SD, USA}
\address{$^{12}$Department of Physics and Astronomy, University of Tennessee, Knoxville, TN, USA}
\address{$^{13}$Research Center for Nuclear Physics and Department of Physics, Osaka University, Ibaraki, Osaka, Japan}
\address{$^{14}$Department of Physics and Astronomy, University of North Carolina, Chapel Hill, NC, USA}
\address{$^{15}$Pacific Northwest National Laboratory, Richland, WA, USA}
\address{$^{16}$Department of Physics, Black Hills State University, Spearfish, SD, USA}
\address{$^{17}$Tennessee Tech University, Cookeville, TN, USA}

\ead{wxu@lanl.gov}

\begin{abstract}
Neutrinoless double-beta (\BBz) decay is a hypothesized process where in some even-even nuclei it might be possible for two neutrons to simultaneously decay into two protons and two electrons without emitting neutrinos. This is possible only if neutrinos are Majorana particles, \textit{i.e.} fermions that are their own antiparticles. Neutrinos being Majorana particles would explicitly violate lepton number conservation, and might play a role in the matter-antimatter asymmetry in the universe. The observation of neutrinoless double-beta decay would also provide complementary information related to neutrino masses. The \textsc{Majorana} Collaboration is constructing the \textsc{Majorana} \textsc{Demonstrator}, with a total of 40-kg Germanium detectors, to search for the \BBz~decay of \ge\ and to demonstrate a background rate at or below 3 counts/(ROI$\cdot$t$\cdot$y) in the 4 keV region of interest (ROI) around the 2039 keV Q-value for \ge\ \BBz\ decay.  In this paper, we discuss the physics of neutrinoless double beta decay and then focus on the \textsc{Majorana} \textsc{Demonstrator}, including its design and approach to achieve ultra-low backgrounds and the status of the experiment. 
\end{abstract}

\section{Introduction}
Since its discovery and confirmation by the Super-Kamiokande~\cite{SuperK}, the Sudbury Neutrino Observatory (SNO)~\cite{SNO} and other experiments~\cite{PDG}, a non-zero neutrino mass has been recognized as an indication of physics beyond the Standard Model of particle physics, and therefore one of the most fundamental questions. Majorana particles are fermions that are their own anti-particles. If neutrinos are Majorana particles, neutrinoless double-beta (\BBz) decay can take place in some even-even nuclei where single-beta decay is energetically forbidden or highly suppressed. In this process, two neutrons simultaneously decay into two protons and two electrons without emitting neutrinos. This would explicitly violate the lepton number conservation~\cite{BBz1, BBz2, BBz3}. \BBz~decay experiments are the only way to unambiguously establish the Majorana or Dirac nature of standard model neutrinos and the observation of \BBz~decay would have a profound impact.\\

\BBz~decay requires neutrinos to be massive and their own antiparticles. A direct relationship can be established between \BBz~decay half time and the effective Majorana neutrino mass as 
 \begin{equation}
(T^{0\nu}_{1/2})^{-1} = G^{0\nu} |M_{0\nu} |^2 \left(\frac{<m_{\beta\beta}>}{m_e}\right)^2,
\end{equation}
where $G^{0\nu}$ is the phase space of the decay, $M_{0\nu}$ is the nuclear matrix element and $<m_{\beta\beta}>=|\Sigma^{3}_{i=1}U^2_{ei}m_i|$ is the effective Majorana neutrino mass with $U_{ei}$ being the Pontecorvo-Maki-Nakagawa-Sakata mixing matrix~\cite{AHEP}. 
Therefore, measurement of the \BBz~decay half time is a complementary way to probe the absolute neutrino mass scale. For example, the inverted mass hierarchy has an effective neutrino mass between about 15 meV and 50 meV, and the lower boundary corresponds to a~$^{76}$Ge~\BBz~decay half time longer than $10^{27}$ years. \\

In recent decades, many experiments have searched for \BBz~decay in several isotopes including, {\it e.g.}, \ge, \xe, \te, and others. For recent reviews of \BBz~decay experiments, see for example references~\cite{BBz3, BBz4}. All of these isotopes also undergo two-neutrino double-beta (\BBt) decay with typical half-lives on the order of $10^{19}$ to $10^{21}$ years~\cite{BBt1, BBt2, BBt3, BBz4}. Similar to a single-beta decay spectrum, the sum energy of the two electrons in \BBt~decay is a continuum extending up to the Q-value (\QBB) of the decay. In contrast, due to the absence of neutrinos, the two electrons in \BBz~decay carry almost all of the energy released in the decay, and their total kinetic energy is peaked at the Q-value. Compared to other detection systems, High-Purity Germanium (HPGe) detectors have the critical advantage of excellent energy resolution, which eliminates the otherwise almost irreducible \BBt~decay background and improves the discovery potential due to the clean line signature.\\

\section{The \mj\ \dem}
The \textsc{Majorana} Collaboration is constructing the \textsc{Majorana} \textsc{Demonstrator} (MJD)~\cite{AHEP}~in class-100 clean rooms located on the 4850' underground level at the Sanford Underground Research Facility (SURF)~\cite{SURF}. The collaboration plans to deploy a total of 40-kg Germanium detectors into two modules, with 30~kg of the Ge detectors enriched to 87\% \ge. The \textsc{Demonstrator} aims at demonstrating a path forward to achieve a ultra low background and at showing technical and engineering scalability toward a tonne-scale instrument.\\

\begin{figure}[!htbp]
\begin{minipage}{38pc}
\includegraphics[width=38pc]{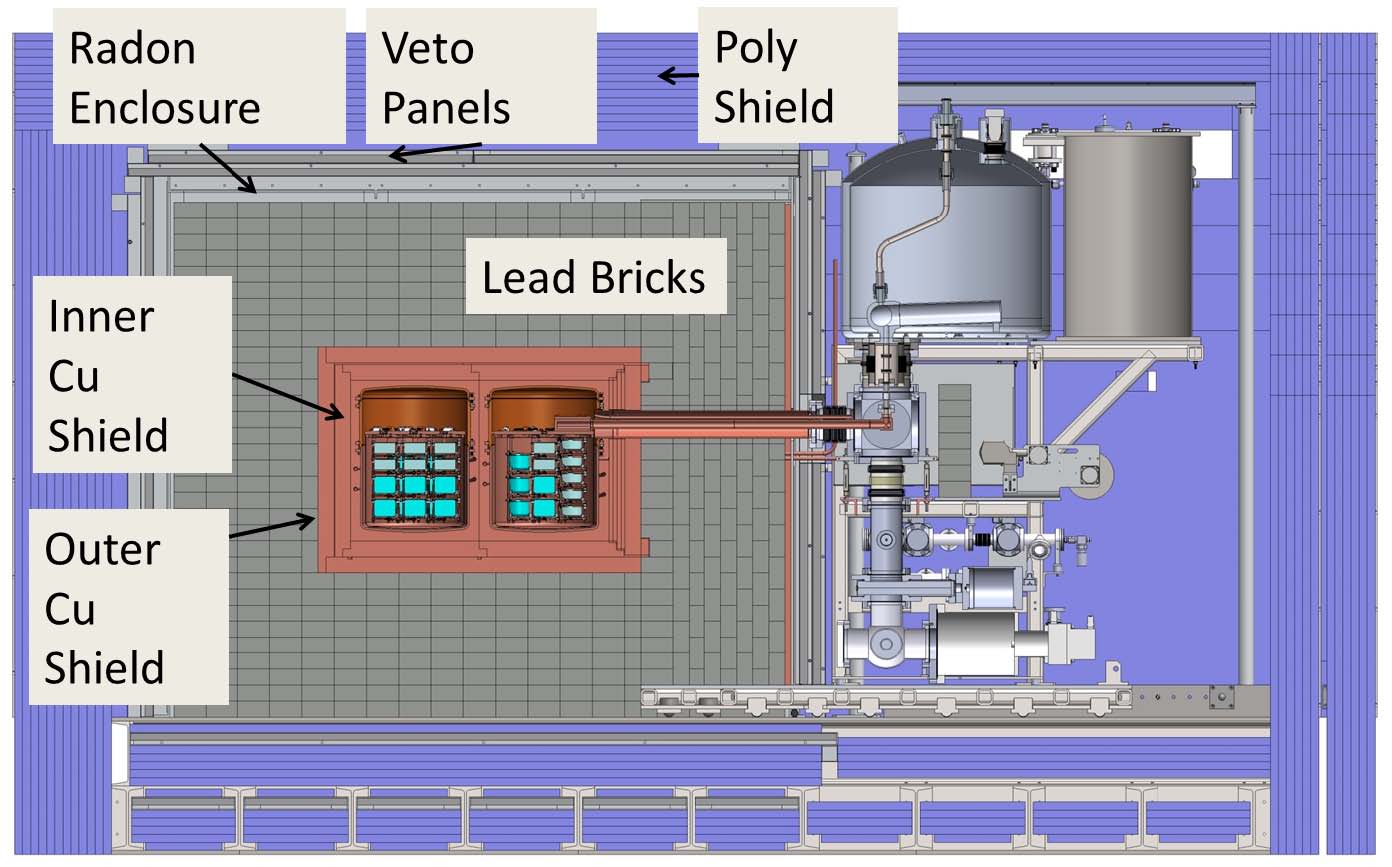}
\end{minipage}
\caption{\label{fig:shield_design} Adapted from~\cite{AHEP}. MJD shield configuration. The experimental apparatus is sitting on the top of a stainless-steel over-floor with orthogonal channels, allowing HDPE panels (poly shield) and veto panels underneath the apparatus. See the text for detailed discussions of the shield.}
\end{figure}

To be specific, a background rate at or below 1.0~count/(ROI$\cdot$t$\cdot$y) in the 4~keV region of interest (ROI) around the 2039~keV Q-value for \ge\ \BBz\ decay is required for tonne-scale Ge-based experiments that will probe the inverted hierarchy parameter space for \BBz\ decay. Based on simulation studies, if the \textsc{Demonstrator} can achieve a background level of 3.0~counts/(ROI$\cdot$t$\cdot$y) or better, then a tonne-scale experiment with similar design and material is expected to have the required background level of 1.0~count/(ROI$\cdot$t$\cdot$y) or better, thanks to various effects such as self-shielding and longer time for cosmogenic backgrounds such as $^{68}$Ge to decay. In addition to searching for \BBz~decay, the \textsc{Demonstrator} will be used to search for a range of new physics beyond the standard model, including low mass dark matter and axions~\cite{AHEP, MALBEK}. \\

A modular approach is chosen for MJD for its natural expandability to larger scale experiment. In the current approach, four or five HPGe detectors are stacked together to form a string, and seven strings are mounted into a single cryostat with dedicated supporting systems such as cryogenic and vacuum systems, together referred to as a module. The construction of the \textsc{Demonstrator} is organized in three phases. First, a prototype module with three strings of natural HPGe detectors was constructed in 2014 and has been in commissioning. By early 2015, the first production module with more than half of the total enriched HPGe detectors and some natural detectors, a total mass of 20~kg of Ge, will be completed. By late 2015, the second production module with the rest of HPGe detectors will be constructed, also with a total mass of 20~kg of Ge. Thanks to separate supporting systems, those modules can be individually operated and maintained, and data-taking will begin as soon as the first module is ready. The rest of this proceedings will report on the status of the MJD project as of September 2014.

\begin{figure}[!htbp]
\begin{minipage}{14.5pc}
\includegraphics[width=14.5pc]{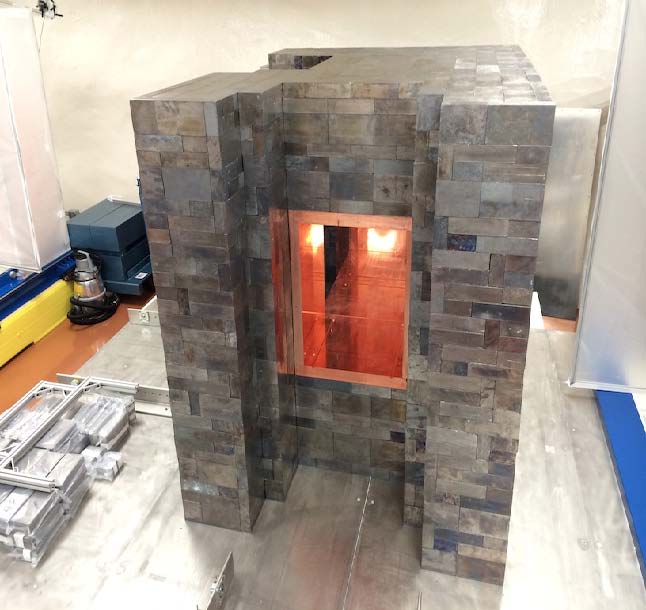}
\end{minipage}\hspace{2pc}%
\begin{minipage}{18pc}
\includegraphics[width=18pc]{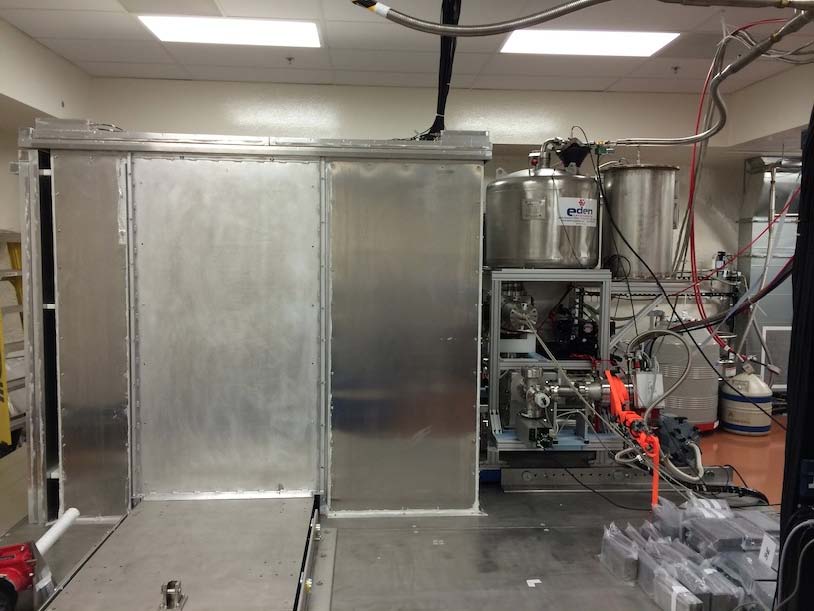}
\end{minipage} 
\caption{\label{fig:shield} Left: A photograph of the main body of the lead shield and outer copper shield. Right: A photograph of the nearly completed shield, inside which the prototype module has been taking commissioning data. The supporting systems of the prototype module can be seen in the photo as well. The cryostat of the prototype module is inside the shield. The current prototype module shielding includes large veto coverage, a semi-sealed Rn box constantly purged with nitrogen gas, complete lead shield at the prototype module side, and outer copper shield surrounding the cryostat. }
\end{figure}

\subsection{Shield}
MJD utilizes both active and passive shielding, as shown in Fig.~\ref{fig:shield_design}. 
The outmost component of the shield configuration is composed of layers of high density polyethylene (HDPE) panels, two layers of which are borated.  The construction of the poly shield has started, and upon completion, it will enclose the entire apparatus, including the supporting systems of each module. Inside the poly shield, there are two layers of active veto panels on every side of the experiment, made of scintillating acrylic sheet and read out by Photomultiplier Tubes (PMTs). The active veto panels on four sides of the shield have been installed and commissioned. 
Inside the veto panels, a semi-sealed Radon (Rn) Enclosure Box made of Aluminum has been constructed to provide a nearly Rn-free environment. A nitrogen gas delivery system that constantly purges the Rn box with boil-off nitrogen gas was implemented. 
Further improvements are currently being implemented to use cold clean charcoal to filter out the remaining Rn in the nitrogen purge gas. Inside the Rn box is a 45 cm thick lead shield composed of $5.1 \times 10.2 \times 20.3$ cm$^3$ lead bricks, which are carefully stacked in a pattern that eliminates any direct path for photons originating from outside the shield to penetrate. The main body of the lead shield has already been completed, as shown in the left side of Fig.~\ref{fig:shield}. In the photo, keyed structures can be seen on two sides of the lead shield. When cryostats are inserted into the shield, additional lead shielding matching the keyed structures will be also installed, completing the lead shield. Inside the lead shield, an outer copper shield made of Oxygen-Free High thermal Conductivity (OFHC) copper has already been installed. An inner copper shield made of electroformed copper will be installed to directly enclose the cryostats. For more details of the shield design, see reference~\cite{AHEP}.  Since August 2014, the MJD prototype module has been taking commissioning data in the nearly completed shield described here, as shown in the photo on the right side of Fig.~\ref{fig:shield}.\\

\subsection{Detectors, Strings, and Modules}
The collaboration has chosen to use P-type Point Contact (P-PC) HPGe detectors~\cite{PPC1, PPC2}. As compared to the more conventional coaxial HPGe detectors, they have several important advantages for a rare-event search. Their small readout contact produces a highly localized ``weighting potential", and this combined with a larger range of charge drift times allows for excellent pulse-shape discrimination sensitivity to distinguish between multi-site and single-site events. This suppresses the main background of Compton-scattered gamma rays. Both P-PC and p-type coaxial detectors also have a thick outer lithium contact that fully absorbs alpha particles, further reducing the background. Lastly, the small capacitance of PPC detectors greatly improves the energy resolution at low energies, reducing the energy threshold and allowing for low-mass dark matter searches.\\
\begin{figure}[!htbp]
\begin{minipage}{11pc}
\includegraphics[width=11pc]{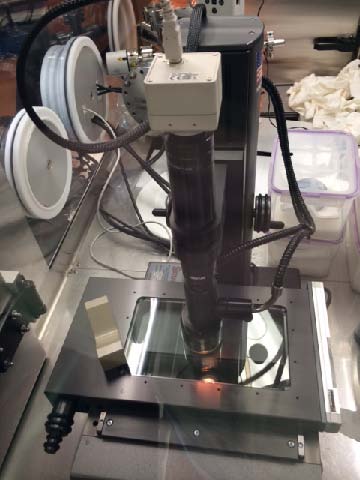}
\end{minipage}\hspace{2pc}%
\begin{minipage}{20pc}
\includegraphics[width=20pc]{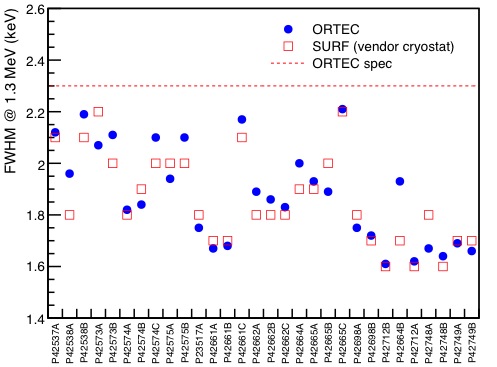}
\end{minipage} 
\caption{\label{fig:Ge}Left: A photograph of a digital microscope that has been used to measure the dimensions of bare Ge crystals inside a custom glove box. A digital scale is used for mass measurements. Right: The Full-Width-at-Half-Maximum (FWHM) at 1333~keV of the 30 enriched ORTEC detectors measured by ORTEC (blue dots) and by the MJD collaboration at SURF (red open squares), plotted against detector serial number. The FWHM of all detectors are better than the experimental specification of 2.3~keV, which is shown as the dotted horizontal line.}
\end{figure}

\begin{figure}[!htbp]
\begin{minipage}{18pc}
\includegraphics[width=18pc]{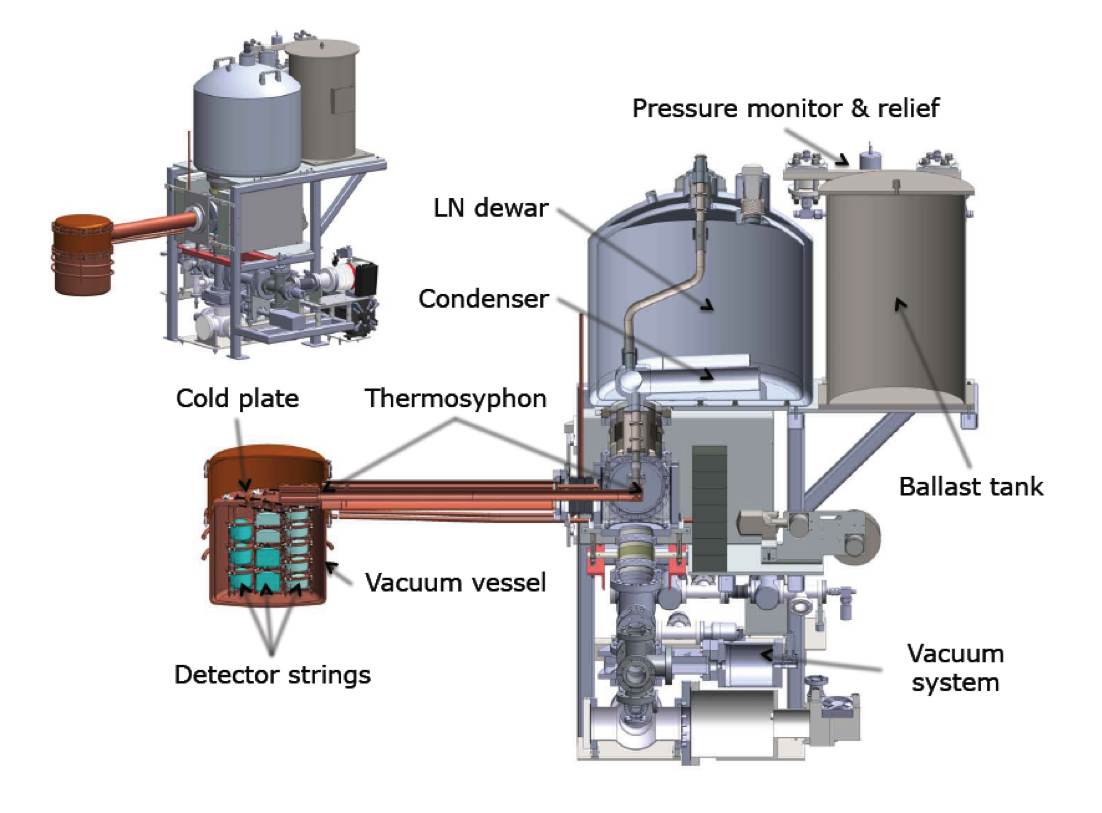}
\end{minipage}\hspace{2pc}%
\begin{minipage}{18pc}
\includegraphics[width=18pc]{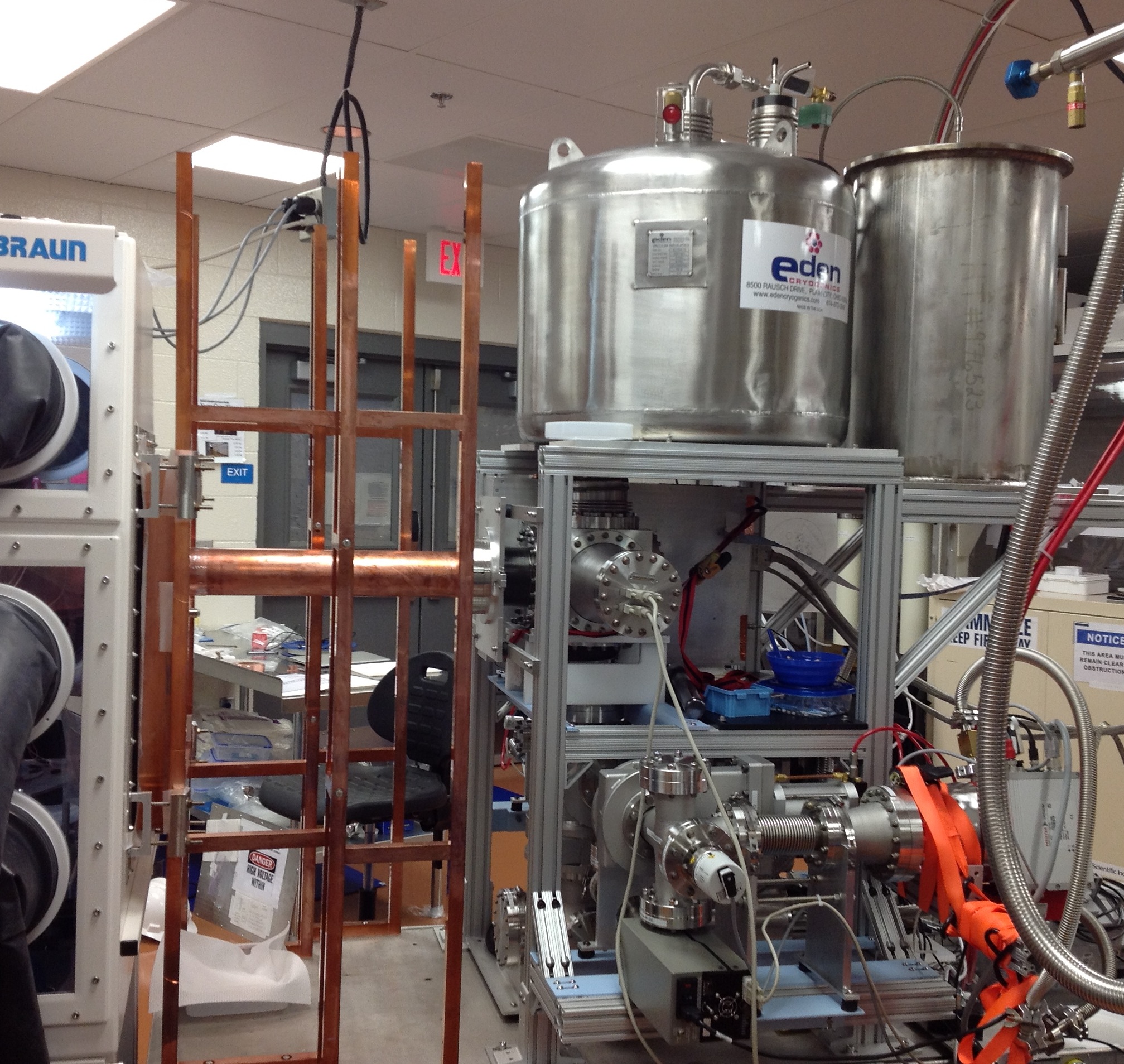}
\end{minipage} 
\caption{\label{fig:module}Left: Adapted from~\cite{AHEP}. The design of a \textsc{Majorana} \textsc{Demonstrator} module. Right: A photograph of the prototype module during string mounting. The unseen cryostat is inside a custom glove box so that it can be opened up to allow string mounting. The visible copper tube that connects the cryostat and the vacuum system is called ``cross-arm tube", within which is the thermosyphon tube. The thermosyphon tube is filled with nitrogen in dual phases. The copper structure surrounding the cross-arm tube is used to support part of the lead shielding, which completes the main body of the lead shield on the over-floor when the cryostat is inserted into the shield.}
\end{figure}

Most of the natural Ge detectors being deployed at MJD are the Broad Energy Ge detectors (BEGe) manufactured by Canberra~\cite{Canberra}, and all of the enriched Ge detectors and two natural Ge detectors are manufactured by AMETEK/ORTEC~\cite{ORTEC}. A comprehensive set of acceptance and characterization tests of the ORTEC detectors have been carried out at SURF~\cite{DetAccp}, including measurements of the masses and dimensions of bare Ge crystals, as shown in the left side photo of Fig.~\ref{fig:Ge}. So far, 30 enriched detectors with a total mass of 25.2 kg have been delivered to SURF and all of them meet experimental specifications such as energy resolution, as shown in the right panel of Fig.~\ref{fig:Ge}. Natural BEGe detectors have also been tested. Efforts to manufacture more detectors to reach 30 kg are still ongoing.\\ 

Bare Ge crystals are removed from their commercial cryostats and then built into detector units packed with custom high voltage contacts and front-end signal read-out boards. Depending on the size of the bare crystals, four or five detector units are stacked into one string, and each string is then tested for performance. The metal supporting parts in the detector units and strings are made of ultra pure electroformed copper~\cite{EForm} produced underground to minimize radioactivity at the vicinity of detectors. More details about detector units, strings and their performance can be found in other MJD talks in this workshop: I.~Guinn,~Low Background Signal Electronics for the \textsc{Majorana} \textsc{Demonstrator}; J.~Leon,~Prototype for a low-noise charge preamplifier with a forward-biased JFET; S.~Mertens,~\textsc{Majorana} Experience with Germanium Detectors; and B.~Jasinski,~Assembly of MJD.\\ 

A total of fourteen production strings of HPGe detectors will be constructed, seven for each production module. The construction of strings for the first module is well underway as of December 2014. In addition, three strings with a total of ten natural HPGe detectors of both BEGe and ORTEC types have been constructed for the prototype module and are being used for commissioning of the experimental apparatus. \\

Each MJD module consists of a cryostat with detector strings, cryogenic and vacuum systems, parts of the shield and its own calibration system, as shown in the design plot in the left panel of Fig.~\ref{fig:module}. Each module can be moved as a whole inside the clean room. 
The cryogenic system is a thermosyphon filled with nitrogen in both gas and liquid phases. The thermosyphon thermally connects to a cold plate inside the cryostat at one end and it is constantly cooled inside a Liquid Nitrogen (LN) dewar at the other end~\cite{thermosyphon}. For production modules, the cryostat itself and the thermosyphon are all made of electroformed copper. \\

\begin{figure}[h]
\begin{minipage}{10pc}
\includegraphics[width=10pc]{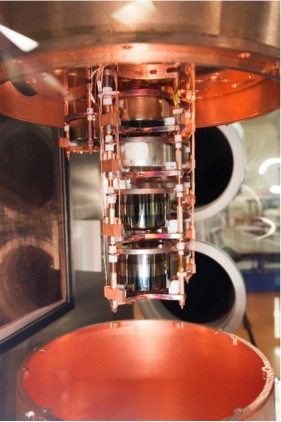}
\end{minipage}\hspace{2pc}%
\begin{minipage}{26pc}
\includegraphics[width=26pc]{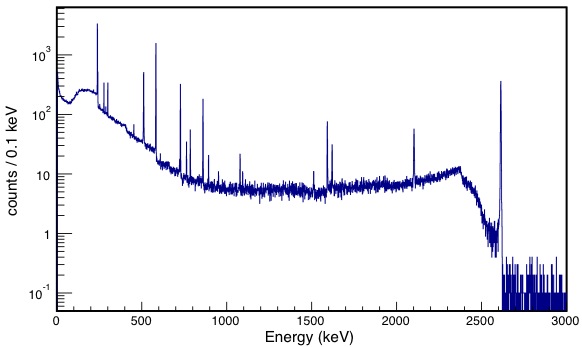}
\end{minipage}
\caption{\label{fig:prototype}Left: A photograph inside the prototype cryostat,  which is opened up to allow the mounting of three strings of HPGe detectors. The string in the front has two Canberra BEGe detectors on the top and two ORTEC detectors at the bottom. Right: A spectrum taken by one of the HPGe detectors in the prototype cryostat during in a calibration run with a $^{228}$Th line source.} 
\end{figure}

To validate the apparatus design and to debug issues in the construction and operation of modules and the DAQ systems, a prototype module was constructed and has been in commissioning, as shown in the right panel of Fig.~\ref{fig:module}. The prototype module is identical to production modules, except it is made of commercial OFHC copper. The vacuum system of the prototype module will be reused for the second production module. Ten natural HPGe detectors of both BEGe and ORTEC types are arranged into three strings and mounted onto the cold plate inside the prototype module cryostat, as shown in the left panel of Fig.~\ref{fig:prototype}. \\

DAQ systems based on the Object-Oriented Real-time Control and Acquisition (ORCA) platform~\cite{ORCA} have been instrumented to control both commercial and collaboration-manufactured electronics, and background data and calibration data are being taken. The HPGe detectors in the prototype cryostat have shown good energy resolution similar to that in commercial cryostats. The experience of commissioning the prototype module has been critical for the collaboration to improve the experimental apparatus. In addition, a set of detailed construction procedures and strict cleanness protocols, including clean machining, have been developed and improved during this process. The prototype module commissioning will continue for a few months and the construction of the first production module is well underway and will be completed in early 2015.\\

\subsection{Background Estimation}
\begin{figure}[hc]
\begin{minipage}{30pc}
\includegraphics[width=30pc]{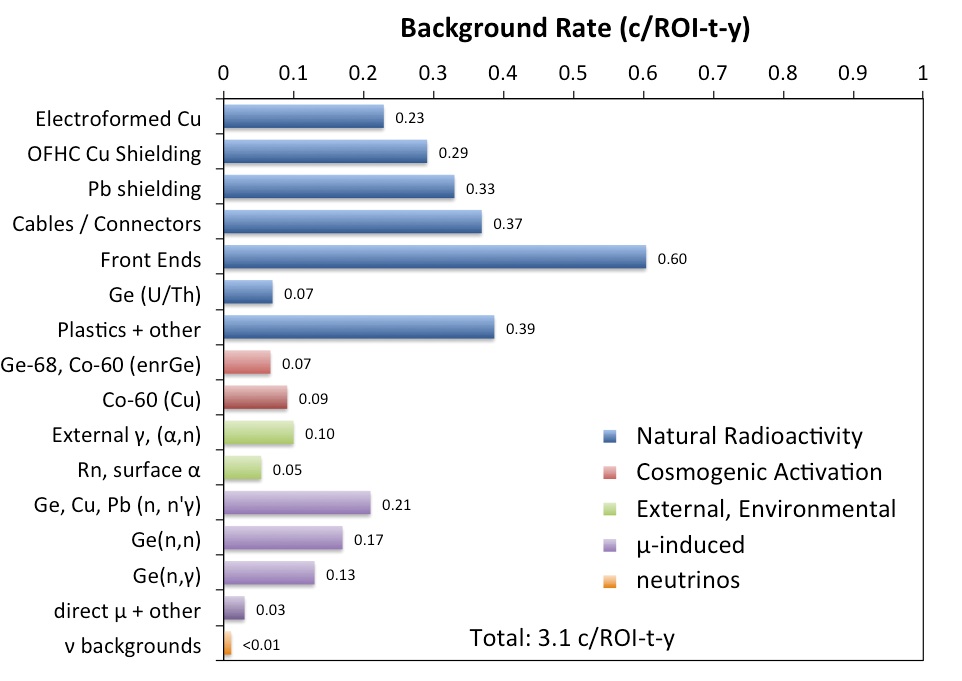}
\end{minipage}
\caption{\label{fig:summary} Estimated background contributions in the \BBz~decay ROI at MJD. The estimations come from simulation studies based on material assay results as well as existing measurements of various physics processes, and some of the estimations are upper limits. The contributions sum to $\le$ 3.1 counts/(ROI-t-y) in the \textsc{Demonstrator}.}
\end{figure}

A vigorous program to assay the natural radioactivity in almost all materials used in the MJD experiment has provided a foundation for accurate background projections, summarized in Fig.~\ref{fig:summary}. The radioactivity of components close to the active HPGe detector volumes has a strong influence on the background level. In particular, ultra pure copper electroformed underground has both much reduced primordial radioactivity and limited cosmogenically produced $^{60}$Co. By using electroformed copper at some locations closest to the detectors, such as the support structures in detector units and strings as well as the cryostat itself, we have achieved several orders-of-magnitude background reduction over the use of commercial alternatives. The biggest projected background is from natural uranium and thorium in the electronics front-end, cables and connectors, which are also very close to the detectors. Some of the estimations are upper limits and work to improve them are still in progress. Summarizing all the contributions, the projected background for MJD is $\le$ 3.1~counts/(ROI-t-y) in the 4~keV ROI around 2039~keV. More details about the MJD background model can be found in references~\cite{AHEP, bk}. 

\section{Summary}
The assembling and construction of the \textsc{Majorana} \textsc{Demonstrator} is proceeding at the 4850' level of the Sanford Underground Research Facility. The shield is near completion and a prototype module with ten natural HPGe detectors was constructed and has been in commissioning. The first production module with 20-kg HPGe P-PC detectors, mostly enriched in \ge\, will be completed and enter commissioning in early 2015; the second production module is expected to be completed in late 2015.

\section{Acknowledgments}
This material is based upon work supported by the U.S. Department of Energy, Office of Science, Office of Nuclear Physics. We acknowledge support from the Particle Astrophysics Program of the National Science Foundation. This research uses these US DOE Office of Science User Facilities: the National Energy Research Scientific Computing Center and the Oak Ridge Leadership Computing Facility. We acknowledge support from the Russian Foundation for Basic Research. We thank our hosts and colleagues at the Sanford Underground Research Facility for their support.

\end{document}